\documentstyle[11pt,moriond,epsfig]{article}

\def\bx{{\bf x}}
\def\by{{\bf y}}
\def\bk{{\bf k}}

\def\half{{\frac{1}{2}}}
\def\be{\begin{equation}}
\def\ee{\end{equation}}
\def\bea{\begin{eqnarray}}
\def\eea{\end{eqnarray}}

\begin{document}
\vspace*{4cm}
\title{Consilience of High-T$_c$ Theories}

\author{J. B. Marston}

\address{Department of Physics, Brown University, Providence, Rhode
Island 02912-1843 USA}

\maketitle\abstracts{Improvements both in the quality and in the variety of
experiments on high-temperature superconductors have yielded new insights 
into the microscopic origins of pairing.  A number of competing theories
have already been ruled out.  Some of the more promising descriptions --
gauge theories, coupled-chains, nesting instabilities, nodal liquids,
and stripes -- share features in common.  A unified picture of the cuprates
is beginning to emerge.}

\section{Introduction}
\label{sec:intro}

In his recent book {\it Consilience: The Unity of Knowledge}, biologist 
E. O. Wilson reintroduced the word ``consilience'' into the English
language.  William Whewell wrote in 1840 that
``The Consilience of Induction takes place when an
Induction, obtained from one class of facts, coincides with an Induction,
obtained from another different class.  This Consilience is a test of the
truth of the Theory in which it occurs~\cite{EOW}.''  The cuprate high
temperature superconductors are complex enough that it is appropriate to apply
this truth principle to sort out various models.  In this brief
overview, I focus on {\it some} of the theories which have 
at least a chance of surviving.  I will argue that recent years have seen a
consilience, a coming-together, of theories of high temperature 
superconductivity.

This lecture was presented in Hanoi at the {\it International Workshop on 
Superconductivity, Magneto-Resistive Materials, and Strongly Correlated 
Quantum Systems}.  Vietnam is a country with a long tradition of 
Buddhism.  It is therefore fitting to recall the second of 
Buddhism's Four Noble Truths:
Attachment leads to suffering.  It is important to not become too attached to 
any one theory of high-temperature superconductivity!  
Some are completely wrong; others mix strong and weak features, and none are
as yet definitive.

\section{Key Experiments}
\label{sec:key}

Recent experimental advances show that the various high temperature 
superconductors share much in common.  
A good overview of the cuprate materials and their key experimental
properties can be found in a recent review by Maple~\cite{Maple}.  
Experiments now agree that there is d$_{x^2-y^2}$ superconducting 
order~\cite{d-wave}, at least in the hole-doped compounds.  One of
the most intriguing phenomena is the existence of a pseudogap in the
density of states at temperatures
{\it above} the superconducting transition temperature $T_c$.  Many 
experiments show, or are at least consistent with, the existence of the 
pseudogap~\cite{Service,Deutscher,Corson,Markiewicz,Kusko}.  
Four other key experimental results are:

1. Rough Electron-Hole Symmetry:
As emphasized by Maple~\cite{Maple}, the behavior of the electron doped 
materials, such as Nd$_{2-x}$Ce$_x$CuO$_{4-y}$, is qualitatively similar
to that of the hole-doped compounds such as La$_{2-x}$Sr$_x$CuO$_{4-y}$.
In particular, both are antiferromagnetic insulators at small $x$ which 
become superconducting at larger values of $x$.  Whether or not the electron
doped materials are d-wave superconductors is a crucial question.  If they
are s-wave, much of the subsequent discussion in this article is falsified, 
or at the very least must be reworked. 

2. Angle Resolved Photoemission (ARPES):
Although ARPES experiments continue to contradict each 
other~\cite{Norman,Ino,Chuang}, some features stand out.  Nesting of the Fermi
surface seems to have been observed~\cite{Dessau}.  And one insulating 
antiferromagnetic parent compound shows a very interesting electron
dispersion~\cite{Ronning}, similar to that of a d-wave superconductor, 
despite the absence of superconductivity, see Sec.~ \ref{sec:mean-field}
below.  

3. Neutron Scattering:
Neutrons are sensitive to both static and fluctuating magnetic moments.  
A useful summary is presented by Mason~\cite{Mason}.  Incommensurate 
magnetic fluctuations have been observed for some time in doped 214 
compounds~\cite{Aeppli,Wells,Lee,Wakimoto}, 
and the deviation of the peak wavevectors from the 
N\'eel ordering wavevector $\vec{Q} = (\pi, \pi)$ is proportional to the
doping.  The recent observation of similar incommensurate peaks in 
the 123 material~\cite{Mook} suggests that this behavior is common, and 
perhaps universal, to the cuprates. 

4. Phase Separation and Stripes:
Although phase separation into {\it static} stripes may occur only in some
of the cuprates~\cite{Tranquada,Julien}, and at special dopings 
such as $x = 1/8$, there is 
mounting evidence for ubiquitous, but slowly shifting stripes, 
in particular from $^{63}$Cu NQR measurements~\cite{Hunt}.  Even when there is
no true long-range stripe order, the stripe dynamics may be slow enough that
the stripes can be treated for most purposes as static.

\section{Hubbard and Related Models}
\label{sec:models}

Anderson's early insight was that the Hubbard model is a good starting point
for theories of cuprate superconductivity~\cite{PWAScience}.  
The one-band model, 
\be
H = -t~ \sum_{<\bx,\by>}~ (c^{\dagger \alpha}_\bx c_{\by \alpha} + H.c.)
+ U~ \sum_\bx~ (c^{\dagger \alpha}_\bx c_{\bx \alpha} - 1)^2\ ,
\ee
can be 
justified starting from the more complete three-band model which describes
the p$_{x,y}$ oxygen orbitals as well as the d$_{x^2 - y^2}$ orbitals of the
copper atoms~\cite{Auerbach}.   
For simplicity only nearest-neighbor hopping, and on-site Coulomb repulsion,
appear in the above Hamiltonian; however, next-nearest neighbor 
hopping can be important.  Furthermore, the assumption that the inverse-square
Coulomb interaction is screened down to a pure on-site repulsion certainly
breaks down if there is phase separation (see Sec. ~\ref{sec:phase-sep}).  

At half-filling, and on a square lattice, nesting instabilities open up a
charge gap and the system becomes an antiferromagnetic insulator.  The low
energy spin degrees of freedom can be described in terms of the electron
operators as:
\be
\vec{S}_\bx = \half c_\bx^{\dag \alpha}~ \vec{\sigma}_\alpha^\beta~
c_{\bx \beta} 
\label{fermi-spin}
\ee
subject to the constraint
\be
n_\bx \equiv c_\bx^{\dag \alpha} c_{\bx \alpha} = 1 \ .
\ee
The representation of the spins in terms of the underlying fermions
is not as economical as the representation of the spins themselves,
since the spin operators of Eq. ~\ref{fermi-spin} are invariant under
$U(1)$ gauge transformations at each lattice point~\cite{Baskaran,AM}
\bea
c_{\bx \alpha} &\rightarrow& e^{i \theta_\bx}~ c_{\bx \alpha}
\nonumber \\
c^{\dag \alpha}_\bx &\rightarrow& e^{-i \theta_\bx}~ c^{\dag \alpha}_\bx
\label{xU(1)}
\eea
because the local phase rotation $\theta_\bx$ cancels out.
We denote this infinite gauge symmetry $U(1)^\infty_x$.
The physical meaning of this local symmetry is simply that the charge
degrees of freedom are frozen out and play no role in the insulating magnet.
The underlying fermions have both charge and spin degrees of freedom, but
as the number of fermions on each site is fixed to be one, only the spin
degree of freedom is active.

Conversely, in the $U \rightarrow 0$ limit, the Hubbard model reduces to 
a non-interacting tight-binding model
\be
H = -t~ \sum_{<\bx,\by>} c_\bx^{\dag \alpha} c_{\by \alpha} + H.c.
\ee
which is clearly not invariant under the $U(1)^\infty_x$ transformations,
Eq. ~\ref{xU(1)}.  Distinct $U(1)$ rotation angles at neighboring sites,
$\theta_x$ and $\theta_y$ do not cancel out.  Physically this just means
that the electron number is not conserved as the electrons hop from site to
site.  However, in momentum space, 
\be
H = \sum_{\bk} \epsilon_{\bk}~ c^{\dag \alpha}_{\bk} c_{\bk \alpha}; 
\ \ \ \ \epsilon_{\bk} = -2 t [\cos(k_x) + \cos(k_y)]\ .
\ee
The Hamiltonian is instead invariant under a different infinity of $U(1)$ 
rotations~\cite{Haldane}, 
now at each point in momentum space, rather than position space:
\bea
c_{\bk \alpha} &\rightarrow& e^{i \theta_\bk}~ c_{\bk \alpha}\ .
\nonumber \\
c^{\dag \alpha}_{\bk} &\rightarrow& e^{-i \theta_\bk}~ c^{\dag \alpha}_{\bk}\ .
\label{kU(1)}
\eea
This $U(1)^\infty_k$ symmetry, like the $U(1)^\infty_x$ symmetry explored 
above, has a clear physical interpretation:
in the absence of interactions and disorder, momentum is a good quantum
number for the single-particle states, as these are infinitely long-lived.
In summary,
\bea
U(1)^\infty_x &\leftrightarrow& {\rm insulating~ solid}
\nonumber \\
U(1)^\infty_k &\leftrightarrow& {\rm conducting~ liquid.}
\eea
The real cuprates lie somewhere between these two extreme limits, and it is
the tension between these two limits that leads to much interesting physics.

Further progress can be made by noting that $U \gg t$; hence, double-occupied
sites are energetically disfavored in hole-doped compounds.  We can enforce
this reduction of the on-site Hilbert space from 4 states to 3 by employing
slave bosons to represent the holes and write the t-J model~\cite{Auerbach}
as follows: 
\be
H = -t~ \sum_{<\bx,\by>}~ (c^{\dagger \alpha}_\bx b_\bx 
c_{\by \alpha} b^\dagger_\by + H.c.)
+ J~ \sum_{<\bx,\by>}~ (\vec{S}_\bx \cdot \vec{S}_\by 
- \frac{1}{4} n_\bx n_\by) \ .
\ee
Every time an electron $c$ hops from site $\bx$ to $\by$, a hole $b$ hops
in the reverse direction.  The holonomic constraint 
$c^{\dagger \alpha}_\bx c_{\bx \alpha} + b^\dagger_\bx b_\bx = 1$ then forces
the total electron number on each site to be equal to zero or one, but not two.

There are many approximate solutions of the t-J model.  In the following 
sections I describe several different starting points which, surprisingly, lead
to rather similar conclusions.  

\section{Mean-Field and Gauge Theories of the Undoped Antiferromagnet}
\label{sec:mean-field}

The first starting point I consider is a systematic solution of the 
$t-J$ model by means of a $1/N$ expansion~\cite{AM,MA}.  The basic idea is to
generalize the usual two types of spin, up and down, to $N$ types, and 
solve the resulting $SU(N)$ t-J model in the $N \rightarrow \infty$ limit.   
Fluctuations disappear in the $N \rightarrow \infty$ limit and complex-valued
mean-fields along the bonds acquire expectation values:
\be
\chi_{\bx \by} = \frac{J}{N} c^{\dagger \alpha}_\bx c_{\by \alpha} \ .
\ee
An additional set of scalar fields, $\phi_\bx(t)$, are introduced as 
Lagrange multipliers to enforce the holonomic constraint on the occupancy.   
Under the local $U(1)$ gauge transformations, Eq. ~\ref{xU(1)}, 
the $\chi$ fields transform like the spatial link variables of a compact 
lattice gauge theory, while the $\phi$ fields transform as the time-component
of the gauge fields:
\bea
\chi_{\bx \by}(t) &\rightarrow& e^{i[\theta_\by(t) - \theta_\bx(t)]}~ 
\chi_{\bx \by}(t)
\nonumber \\
\phi_\bx(t) &\rightarrow& \phi_\bx(t) + \frac{d \theta_\bx(t)}{dt}\ .
\eea

One annoying problem with the large-N limit is that, close to half-filling,
the lowest-energy solution is dimerized: $\chi_{\bx \by}$ is non-zero
only on disconnected bonds.  This solution, which is unphysical, can be 
eliminated by the introduction of biquadratic spin-spin interactions~\cite{MA} 
which do nothing to the original $SU(2)$ model (as they reduce to the usual
bilinear exchange at $N=2$) but which suppress dimerization for $N > 2$.
Furthermore, instantons do not induce dimerization as they do in the bosonic
$SU(N)$ and $Sp(N)$ formulations~\cite{absence}.
This is clearly demonstrated by an exact solution which shows no
dimerization~\cite{C-break}.   

\begin{figure}
\epsfig{figure=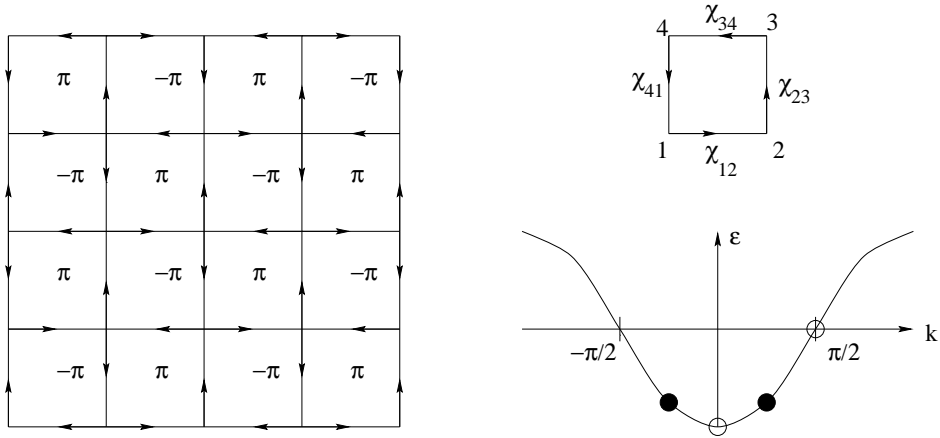,height=2.5in}
\caption{Orientation of the $\chi_{\bx \by}$ fields in the $\pi$-flux phase.
Since $-\pi$ flux is equivalent to $\pi$ flux, this phase has the full
translational symmetry of the underlying lattice and respects time-reversal
symmetry.  By considering a single
plaquette of just four sites, it is clear that $\pm \pi$ flux is energetically 
favored 
as the filled electronic states have lower net energy (solid dots) compared
to zero flux (open dots).    
\label{fig:flux}}
\end{figure}

\begin{figure}
\epsfig{figure=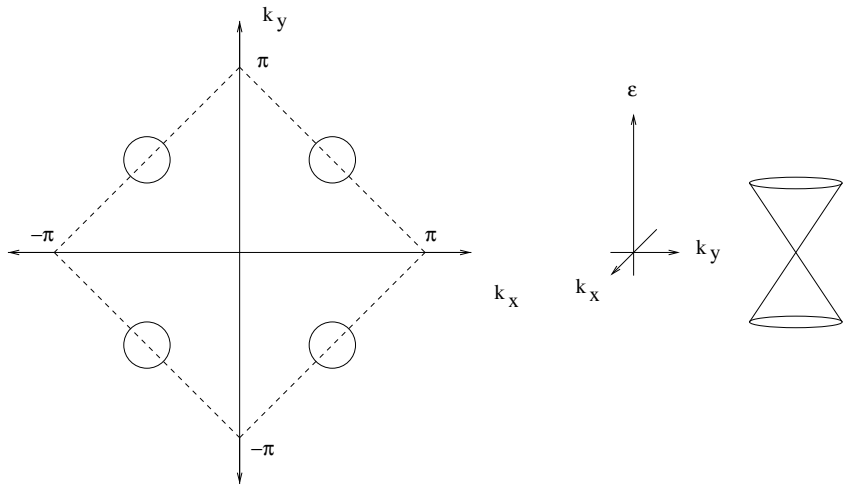,height=2.5in}
\caption{Dispersion of the flux phase.  There is a pseudo-gap everywhere 
except at the nodal points $\vec{k} = (\pm \pi/2, \pm \pi/2)$. 
These points are paired and the
dispersion is linear (cone) like that of massless Dirac fermions.
\label{fig:dispersion}}
\end{figure}

At half-filling the ground state is the $\pi$-flux phase~\cite{AM,Kotliar}.  
The flux is defined by the gauge invariant plaquette operator  
$\langle \chi_{12} \chi_{23} \chi_{34} \chi_{41} \rangle < 0$ with the 
$\chi$ fields oriented as shown in Fig. ~\ref{fig:flux}.  Rokhsar pointed
out~\cite{Rokhsar} a simple way to see that non-zero flux is energetically 
favored: it lowers the kinetic energy of the fermions on the lattice, an 
effect which is particularly easy to see for the four-site problem, 
see Fig. ~\ref{fig:flux}.  
The spectrum is markedly different from that of a tight-binding model as now
the fermion dispersion has the form:
\be
\epsilon(\vec{k}) \propto \pm \sqrt{\cos^2(k_x) + \cos^2(k_y)}\ .
\ee    
The negative energy states are filled, and 
the nesting instability has been removed by the appearance of a pseudogap
everywhere except at the discrete points $\vec{k} = (\pm \pi/2, \pm \pi/2)$
where the density of states vanishes linearly as shown in 
Fig. ~\ref{fig:dispersion}.  
It is remarkable that precisely this dispersion is seen in ARPES 
experiments~\cite{Ronning}
on the insulating parent compound Ca$_2$CuO$_2$Cl$_2$, 
an observation emphasized by Laughlin~\cite{Laughlin}. 

Because the mean-fields $\chi$ and $\phi$ are singlets under $SU(N)$ spin
rotations, long-range spin-order is impossible in the $N \rightarrow \infty$
limit.  To recover N\'eel order requires the consideration of non-perturbative
$1/N$ corrections -- a notoriously difficult problem.  
Some progress~\cite{instantons,Kim} 
was made by recognizing that formation of the long-range spin order is 
equivalent to the dynamical generation of fermion mass~\cite{instantons} in 
$2+1$ dimensional $U(1)$ gauge theory, a problem that has been studied by 
particle theorists.  At large-N there is a global $SU(2N)$ symmetry 
which combines the $N$ spin species with the 
two types of gapless points in the reduced Brillouin zone 
at $\vec{k} = (\pm \pi/2, \pi/2)$. This symmetry breaks down at finite-N as 
$SU(2N) \rightarrow SU(N) \otimes SU(N)$
when the fermions become gapped.  Other
massless excitations then arise, which can be identified as the Goldstone
modes or spin-waves of the N\'eel ordered magnetic state, or equivalently as
bound states of particles and holes, otherwise known as mesons.    

An interesting analysis of $1/N$ corrections to the $U \rightarrow \infty$
Hubbard model, with no spin exchange interaction $J$, 
has recently been carried 
out~\cite{Ziqiang}.  At leading order, as expected, the system is a Fermi 
liquid for any doping away from half-filling.  However, by working to order
$1/N$ and then setting $N=2$, Fermi liquid behavior is found to be destabilized
at light doping $x < 0.07$, in rough agreement with the large-N results for
the $t-J$ model.  

\section{Phase Separation at Non-Zero Doping and Incommensurate Order}
\label{sec:phase-sep}

At non-zero doping there is the possibility of a staggered flux phase (SFP). 
In the SFP the hole occupancy is assumed to be uniform, 
$\langle b^\dagger_\bx b_\bx \rangle = N x / 2$, and 
the flux is reduced in magnitude from $\pi$, eventually disappearing altogether
at large enough doping.  
Time-reversal and translational symmetries are broken in the SFP and 
orbital currents with real magnetic fields arise.  These 
appear to be ruled out experimentally~\cite{2-observe}.   

On the other hand,
phase separation into stripes had been predicted on theoretical 
grounds~\cite{stripe}, prior to any clear experimental observation of the
phenomenon.  The stability of stripes depends on the size of the hopping
matrix elements beyond nearest-neighbor exchange~\cite{stripe-stability}. 
Phase separation solves the problem of the breaking of time-reversal and 
translational symmetries in the SFP.  Phase separation into stripes also 
can explain the incommensurate, and nearly critical~\cite{Aeppli}, 
spin fluctuations seen in neutron scattering experiments in the 214 
compound~\cite{Wells} and recently in the 123 material~\cite{Mook}.  The two
problems are solved simultaneously by assuming that the electron-rich region
is an antiferromagnetic insulator described by the $\pi$-flux phase, with its
strong tendencies towards spin ordering.  
The infrared divergences~\cite{instantons} 
at finite-N are cut-off by the limited size of the electron-rich region. 
Incommensurate spin peaks arise due to these finite-size effects.
The tendency to phase separate can also 
be viewed fruitfully~\cite{Rome} in terms
of a Fermi liquid close to a quantum critical point (QCP).

\section{Squares, Chains, and Ladders}
\label{sec:squares}

Another way to study the behavior of the t-J model on the square 
lattice is to build up the lattice systematically from smaller subunits.  
For example, the four site system, a square plaquette, can be easily 
diagonalized.  At half-filling, the spin-spin correlations in the ground
state are such that the expectation value of the plaquette operator has a
negative value.  To be precise, in the physical $N=2$ case, 
the expectation values of three-spin operators such as  
$\langle \vec{S}_1 \cdot (\vec{S}_2 \times \vec{S_3}) \rangle$ vanish if
there is no time-reversal symmetry breaking, and
\bea 
\langle \chi_{12} \chi_{23} \chi_{34} \chi_{41} \rangle  
&=& \frac{J^4}{16}~ \bigg{\{} \frac{1}{8} 
+ \frac{1}{2} [ \langle \vec{S}_2 \cdot \vec{S_3} \rangle 
+ \langle \vec{S}_2 \cdot \vec{S_4} \rangle 
+ \langle \vec{S}_3 \cdot \vec{S_4} \rangle 
- \langle \vec{S}_1 \cdot \vec{S_2} \rangle 
- \langle \vec{S}_1 \cdot \vec{S_3} \rangle 
- \langle \vec{S}_1 \cdot \vec{S_4} \rangle] 
\nonumber \\
&-& 2~ 
[\langle (\vec{S}_1 \cdot \vec{S_2}) (\vec{S}_3 \cdot \vec{S_4}) \rangle  
+ \langle (\vec{S}_1 \cdot \vec{S_4}) (\vec{S}_2 \cdot \vec{S_3}) \rangle  
- \langle (\vec{S}_1 \cdot \vec{S_3}) (\vec{S}_2 \cdot \vec{S_4}) \rangle]
\bigg{\}} < 0\ .  
\eea
Thus flux $\pi$ penetrates the plaquette.  The same
result holds for the $4 \times 4$ lattice with periodic boundary 
conditions~\cite{Runge} providing added support for the flux phase.  

The square lattice can also be built up chain-by-chain.  A single Hubbard
chain, with repulsive interactions, is a Luttinger liquid away from 
half-filling; at half-filling a gap develops in the charge sector, while
the spin sector remains gapless in accord with the physics of a spin-1/2
quantum antiferromagnetic chain.  S-wave Cooper pairing is inhibited by the 
strong on-site repulsion and there can be no d-wave superconducting tendencies
along a single chain. 

\begin{figure}
\epsfig{figure=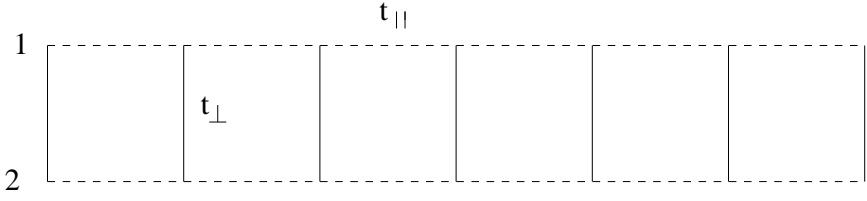,height=1.0in}
\caption{A ladder consisting of two coupled Hubbard chains, numbered 1 and 2,
with intrachain hopping in the x-direction, $t_\parallel$, 
and interchain hopping in the y-direction, $t_\perp$.}
\label{fig:ladder}
\end{figure}

Coupling two such chains with transverse
hopping $t_\perp$, as shown in Fig. ~\ref{fig:ladder}, leads to more 
interesting possibilities.  
Following Fisher~\cite{LesHouches} the first step is to
diagonalize the $U=0$ problem by introducing bonding ($k_y = 0$) 
and anti-bonding ($k_y = \pi$) orbitals:      
\bea
k_y = 0: \ \ \ \
b_{x \alpha} &\equiv& \frac{1}{\sqrt{2}}~ [c_{x \alpha 1} + c_{x \alpha 2}],
\ \ \ \ \epsilon_{k_x} = -2 t_\parallel \cos(k_x) - t_\perp
\nonumber \\
k_y = \pi: \ \ \ \
a_{x \alpha} &\equiv& \frac{1}{\sqrt{2}}~ [c_{x \alpha 1} - c_{x \alpha 2}],
\ \ \ \ \epsilon_{k_x} = -2 t_\parallel \cos(k_x) + t_\perp\ .
\eea
Now if the Hubbard repulsion is turned back on, the RG flows run away to large
values suggesting that charge and spin gaps form and the problem should
be studied from the strong-coupling viewpoint.  
Consider the limit $t_\perp \gg t_\parallel$, as it is
clear that at half-filling spins on opposing sites will pair into singlets
because
$J_\perp \approx 4 t_\perp^2/U \gg J_\parallel \approx 4 t_\parallel^2/U$.
The ground state then may be written:
\be
| \Psi_0 \rangle = 
\prod_x \frac{1}{\sqrt{2}} 
[c^{\dagger \uparrow}_{x 1} c^{\dagger \downarrow}_{x 2} 
- c^{\dagger \downarrow}_{x 1} c^{\dagger \uparrow}_{x 2} ]~ | 0 \rangle
= \prod_x \frac{1}{\sqrt{2}} 
[b^{\dagger \uparrow}_{x} b^{\dagger \downarrow}_{x} 
- a^{\dagger \uparrow}_{x} a^{\dagger \downarrow}_{x} ]~ | 0 \rangle\ .
\ee
Thus, although there is no true off-diagonal long-range order, there is a
d$_{x^2 - y^2}$ wave tendency because the last line looks like a Cooper-paired
state with the sign of the pairing amplitude changing as $k_y$ switches from
$0$ to $\pi$.  

\section{Nodal Liquid}
\label{sec:nodal}
A complementary approach to understanding the underdoped cuprates takes
as its starting point the d-wave superconducting 
state~\cite{nodal,dual-vortex} with dispersion ${\cal E}(\vec{k})$ 
determined, using the tight-binding spectrum 
$\epsilon_{\vec{k}} \propto \cos(k_x) + \cos(k_y)$ 
and d-wave pairing gap function $
\Delta_{\vec{k}} \propto \cos(k_x) - \cos(k_y)$, 
by the usual BCS calculation: 
\be
{\cal E}(\vec{k}) = \pm \sqrt{\epsilon^2_{\vec{k}} +
\Delta^2_{\vec{k}}} \propto \pm \sqrt{\cos^2(k_x) + \cos^2(k_y)}\ .
\ee
This ``nodal liquid'' theory, as its name suggests, focuses on the gapless
nodal points at $\vec{k} = (\pm \pi/2, \pm \pi/2)$. 
As the doping is reduced a zero-temperature quantum phase transition to a 
non-superconducting, but pseudo-gapped, insulating state is envisioned.  
The ``nodons,'' 
or low-energy fermion quasiparticles near the nodes, carry the same quantum
numbers as the fermions in the flux phase~\cite{nodal}.  Upon further 
reduction of the doping, antiferromagnetic order may arise~\cite{Kwon2}.

\section{Isotropic RG Analysis}
\label{sec:RG}

The interplay between nesting, umklapp processes, and the formation of
charge-density-wave (CDW), spin-density-wave (SDW),
and BCS instabilities has been illustrated in a simple model of the Fermi
surface which reduces the nested planes to just four Fermi 
points~\cite{four-points}.
In a recent paper, Furukawa and Rice use a generalized version of this
model to argue that the spin gap in the cuprate superconductors
can be understood as a consequence of umklapp
processes~\cite{Rice}.  However, the whole Fermi
surface can be retained.  High-energy degrees of freedom are integrated out: 
In degenerate fermion systems this means that 
the inner and outer sides of the momentum shell which encloses the
Fermi surface are shrunk successively~\cite{Shankar}.
The RG flow can be investigated by
multidimensional bosonization~\cite{Hyok-Jon,multidimensional}.
The advantage of the bosonized RG calculation is that the existence of
well-defined quasiparticles is not assumed {\it a priori}.  Furthermore,
interaction channels which possess $U(1)^\infty_k$ symmetry
can be diagonalized exactly at the outset.
Similar results have been obtained in the fermion 
basis~\cite{Virosztek,parquet,Kishine,Zanchi}.

\begin{figure}
\epsfig{figure=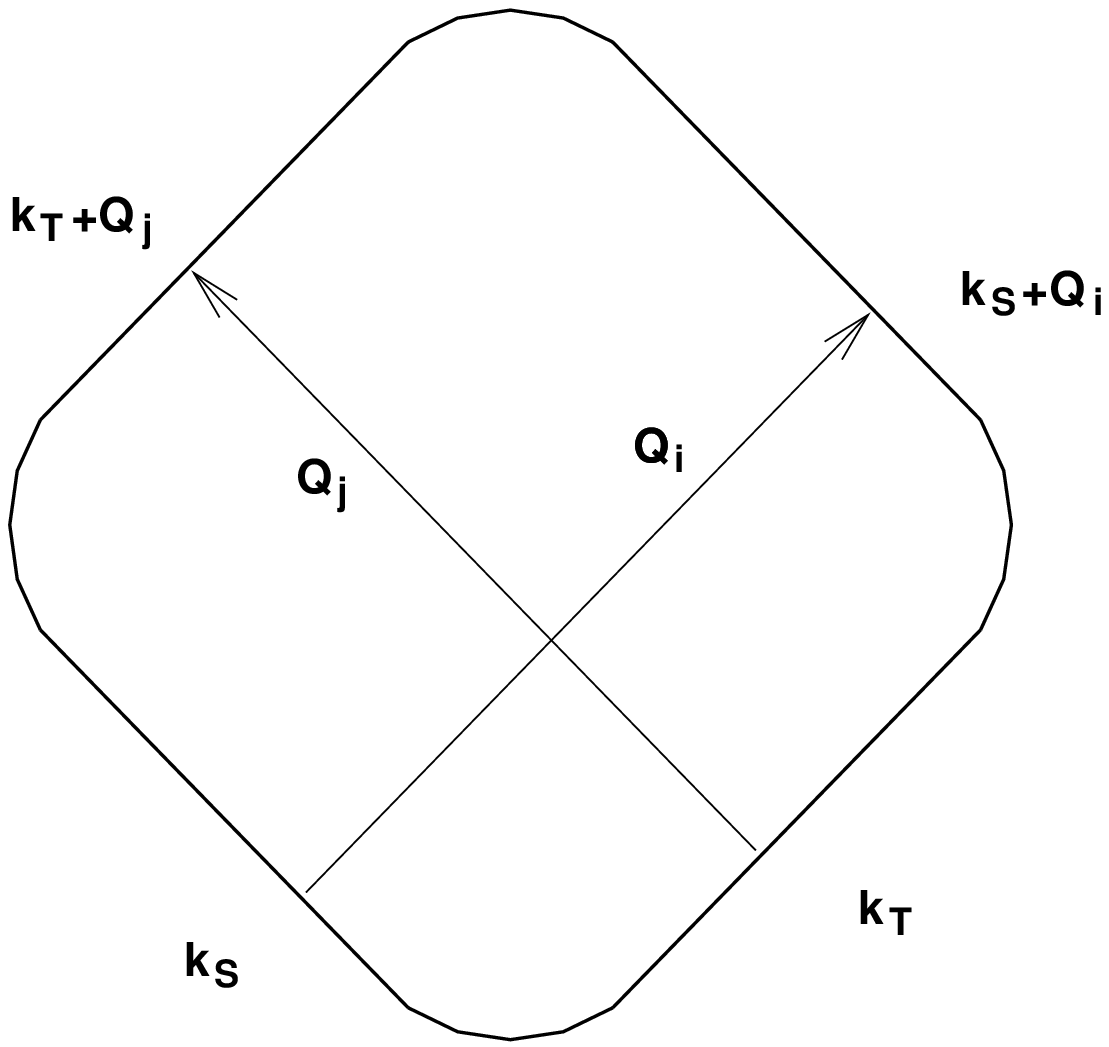,height=2.0in}
\epsfig{figure=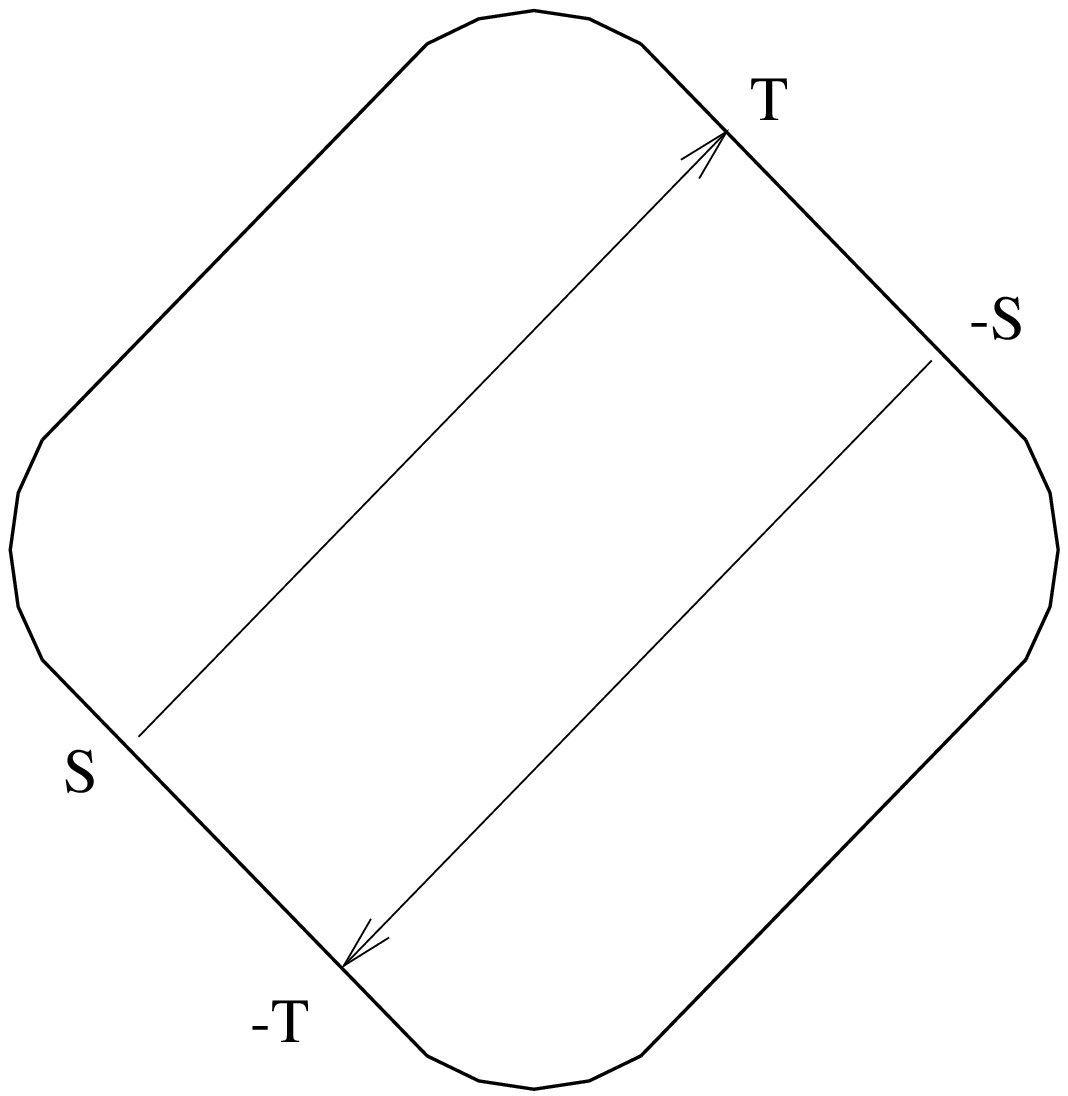,height=2.0in}
\caption{(a) A typical density-wave scattering channel which is permitted
because ${\bf Q}_i + {\bf Q}_j$ is a reciprocal lattice wavevector. 
(b) A typical overlap between density-wave and BCS channels.}
\label{fig:nest}
\end{figure}

Nested Fermi surfaces support many more low-energy
scattering processes than circular Fermi 
surfaces~\cite{multidimensional}.  
Some important processes are depicted in Fig. ~\ref{fig:nest}.
Consequently, the quasiparticle lifetime $\tau \propto 1/T$ is much shorter 
than that in an ordinary Landau
Fermi liquid ($\tau \propto 1/T^2$).  Instead, the normal state is a 
marginal Fermi liquid and the agreement with transport 
data~\cite{MFL} is compelling.  

The tendency towards spin-density order dominates all other
instabilities close to half-filling.  But slightly below half-filling,
because nesting is not perfect but nonetheless effective,
there is competition between the SDW and BCS instabilities.
As the energy cutoff $\epsilon_c$
around the Fermi surface is reduced via the RG
transformation towards $\epsilon_s$, where $\epsilon_s$
is the energy deviation of the actual
Fermi surface from perfect nesting, the SDW channel stops
flowing but the BCS channel continues to renormalize logarithmically
as its flow is independent of nesting.
If the initial SDW coupling is small enough
or if $\epsilon_s$ is large enough (the
SDW channel does not develop an instability
until $\epsilon_c \rightarrow \epsilon_s$) SDW order will not occur.
Instead, d-wave superconductivity sets in at a sufficiently low temperature.

\section{Consilience?}
\label{sec:consilience}

It should be noted that just about any theory which attempts to explain 
cuprate superconductivity in terms of an underlying electronic mechanism will
favor d-wave order~\cite{spin-fluct}.  More revealing is a comparison of 
other features of such theories. 
A summary of the four approaches discussed here, along with the more
phenomenological $SO(5)$ theory of Zhang and collaborators~\cite{so5} is
presented in Table \ref{tab:consilience}.  Noted in the final column are
theories with nodal excitations.
\begin{table}[t]
\caption{Comparison of different theories.\label{tab:consilience}}
\vspace{0.4cm}
\begin{center}
\begin{tabular}{|c|c|c|l|}
\hline
& & & \\
Theory &
Quantum Magnetism &
Proximity to QCP(s) &
$\vec{k} = (\pm \pi/2, \pm \pi/2)$
\\ \hline
Gauge Theory &
$\surd$ &
$\surd$ &
$\surd$ \\
Nodal Liquid &
? &
$\surd$ &
$\surd$ \\
RG Analysis &
$\surd$ &
$\surd$ &
$\surd$ \\
Stripes &
$\surd$ & 
$\surd$ & 
This article\\
SO(5) &
$\surd$ &
$\surd$ &
No \\
\hline
\end{tabular}
\end{center}
\end{table}
It remains to be seen whether or not one unifying description will emerge
from these pieces of the high-T$_c$ puzzle.

\section*{Acknowledgments}
I thank A. Houghton, H.-J. Kwon, C. Nayak and Z. Wang for recent helpful 
discussions.  I would also like to thank the organizers of the 
{\it International Workshop} for their work in bringing the many foreign 
participants to Vietnam.
This research was supported in part by the United States National Science
Foundation under Grants Nos. DMR-9357613 and DMR-9712391.

\end{document}